\documentclass[10pt,fleqn,a4paper,twoside]{article}
\usepackage{abcm}
\def\shortauthor{D. F. Abreu, C. Junqueira-Junior, E. T. V. Dauricio and J. L. F. Azevedo}
\def\shorttitle{A Comparison of Low and High-Order Methods for the Simulation of Supersonic Jet Flows}

\begin{document}
\fphead
\hspace*{-2.5mm}\begin{tabular}{||p{\textwidth}}
\begin{center}
\vspace{-4mm}
\title{COB-2021-0388 \\
A COMPARISON OF LOW AND HIGH-ORDER METHODS FOR THE SIMULATION OF SUPERSONIC JET FLOWS}
\end{center}
\authors{Diego F. Abreu} \\
\institution{Instituto Tecnológico de Aeronáutica, 12228--900, São José dos Campos, SP, Brazil} \\
\institution{mecabreu@yahoo.com.br} \\
\\
\authors{Carlos Junqueira-Junior} \\
\institution{Arts et Métiers Institute of Technology, DynFluid, CNAM, HESAM University, 151 Boulevard de l'Hôpital, 75013, Paris, France} \\
\institution{junior.junqueira@ensam.eu} \\
\\
\authors{Eron T. V. Dauricio} \\
\institution{Instituto Tecnológico de Aeronáutica, 12228--900, São José dos Campos, SP, Brazil} \\ 
\institution{eron.tiago90@gmail.com} \\
\\
\authors{João Luiz F. Azevedo} \\
\institution{Instituto de Aeronáutica e Espaço, 12228--904, São José dos Campos, SP, Brazil} \\
\institution{joaoluiz.azevedo@gmail.com} \\
\\
\abstract{\textbf{Abstract.} The present work compares results for different numerical methods in search of alternatives to improve the quality of large-eddy simulations for the problem of a supersonic turbulent jet flows. Previous work has analyzed supersonic jet flows using a second-order, finite difference solver based on structured meshes, and the results indicated a shorter potential core of the jet and different levels of velocity fluctuations. In the present work, the results of previous simulations are compared to new results using a high-order, discontinuous Galerkin solver for unstructured meshes. All simulations are performed keeping the total number of degrees of freedom constant. The results of the current simulations present very similar mean velocity distributions and slightly smaller velocity fluctuations, and they seem to correlate better with the experimental data. The present results indicate that additional studies should focus on the jet inlet boundary conditions in order to improve the physical representation of the early stages of the jet development.}\\
\\
\keywords{\textbf{Keywords:} Jet flows, Large Eddy Simulation, Discontinuous Galerkin Method}\\
\end{tabular}

\section{INTRODUCTION}

The use of large-eddy simulations (LES) for solving compressible turbulent jets is growing due to the ability to provide sufficiently good solutions at a reasonable cost, when compared to direct numerical simulations (DNS) or the realization of physical experiments. The application of LES on jet flows is interested in obtaining the flow and temperature fields, providing information for aerodynamics, acoustics and heat transfer analyses.

There are many numerical options to perform LES for compressible turbulent flows. Two major distinct approaches are based on low-order methods and high-order methods. When using low-order methods, the good results are obtained from sufficiently fine grids or using adaptive mesh algorithms. The group of high-order methods, instead of depending on fine grids, chooses to work with more sophisticated numerical algorithms for solving the physical equations that allow a more rigorous representation of the equations calculated. Moreover, since high-order methods implicitly increase the number of degrees of freedom, it is possible to have less refined meshes when compared to the low-order methods.

A compressible LES tool, JAZzY, was specifically designed to generate unsteady 
flow data of compressible turbulent jets based on a second-order spatially accurate method for structured 
grids. A validation of JAZzY is performed in \citet{Junior2018} for a supersonic 
jet flow configuration. The validation results indicate a shorter potential core 
of the jet when compared to experimental and numerical data. The potential core 
of the jet is defined as the position in the centerline of the jet where the 
velocity equals 95\% of the jet velocity. Moreover, one can also observe a 
mismatch of the mean velocity profiles along the lipline when comparing to the 
references. The shorter potential core is achieved when using JAZzY with 
different mesh refinements and subgrid-scale (SGS) models.

Structured, low-order methods require very refined meshes near the core of the jet 
and such an approach does not allow local mesh refinement. Therefore, the cost of 
calculations is prohibitive. Once the strategy of using structured grids showed 
some disadvantages for the problem of interest, the authors decided to investigate 
the possibilities using an unstructured, high-order framework, FLEXI 
\citep{Krais2021}, which implements the discontinuous Galerkin (DG) method proposed 
by \citet{Kopriva2010} and \citet{Hindenlang2012}. The FLEXI tool applies a 
discretization that can be employed on unstructured meshes, thus providing more 
flexibility with regard to geometries and meshes. The order of accuracy of the numerical 
algorithm varies by choosing different polynomial degrees for interpolating the 
solution inside each element. The numerical code is open-source and it is already 
validated with interesting results at reasonable costs \citep{Gassner2013}.



The present work is interested in comparing the effects of the DG method and the structured 
2nd-order spatial discretization approach on the solution of supersonic jet flows. 
The comparisons presented here investigate the quality of the results from low-order 
simulations with the two methodologies and the results using FLEXI with third-order 
accuracy. The results are also compared with experimental data. One important aspect 
of the present study is the comparison of different approaches when using similar meshes 
topologies with, also, similar numbers of degrees of freedom in the simulations and, hence, 
comparable computations resources. 
%

\section{NUMERICAL FORMULATION}

\subsection{Governing Equations}
The governing equations to be solved by both numerical methods are the compressible filtered Navier-Stokes equation. In conservative form, they can be expressed by
\begin{equation}
\frac{\partial \mathbf{\bar{Q}}}{\partial t} + \nabla \cdot \mathbf{F} ( \mathbf{\bar{Q}}, \nabla \mathbf{\bar{Q}})=0,
\label{eq.1}
\end{equation}
where $\mathbf{\bar{Q}}=[\bar{\rho}, \bar{\rho} \tilde{u}, \bar{\rho} \tilde{v}, \bar{\rho} \tilde{w}, \bar{\rho} \check{E}]^{T}$ is the vector of filtered conserved variables and $\mathbf{F}$ is the flux vector. The flux vector can be divided into the Euler fluxes and the viscous flux, $\mathbf{F}=\mathbf{F}^e-\mathbf{F}^v$. The fluxes with the filtered variables may be written as
\begin{equation}
\mathbf{F}_i^e= \left[ \begin{array}{c} 
                    \bar{\rho} \tilde{u}_i \\ \bar{\rho} \tilde{u} \tilde{u}_i + \delta_{1i} \bar{p}\\ \bar{\rho} \tilde{v} \tilde{u}_i + \delta_{2i}\bar{p} \\ \bar{\rho} \tilde{w} \tilde{u}_i + \delta_{3i}\bar{p} \\ (\bar{\rho} \check{E} + \bar{p}) \tilde{u}_i
                   \end{array} \right]  \hspace*{1.5 cm} 
\mathbf{F}_i^v= \left[ \begin{array}{c} 
                    0 \\ \tau_{1i}^{mod} \\ \tau_{2i}^{mod} \\ \tau_{3i}^{mod} \\ \tilde{u}_j \tau_{ij}^{mod} - q_i^{mod}
                   \end{array} \right] \hspace*{1.5 cm} \mbox{ , for } i = 1,2,3 ,                  
\end{equation}
where $\tilde{u}_i$ or $(\tilde{u}, \tilde{v}, \tilde{w})$ are the Favre averaged velocity components, $\bar{\rho}$ is the filtered density, $\bar{p}$ is the filtered pressure and $\bar{\rho} \check{E}$ is the filtered total energy per unit volume. The terms $\tau_{ij}^{mod}$ and $q_{i}^{mod}$ are the modified viscous stress tensor and heat flux vector, respectively, and $\delta_{ij}$ is the Kronecker delta. The filtered total energy per unit volume, according to the definition proposed by \citet{Vreman1995} in its "system I", is given by
\begin{equation}
\bar{\rho} \check{E} = \frac{\bar{p}}{\gamma - 1} + \frac{1}{2}\bar{\rho}\tilde{u}_i\tilde{u}_i.
\end{equation}

The filtered pressure, Favre averaged temperature and filtered density are correlated using the ideal gas equation of state $\bar{p}= \bar{\rho} R \tilde{T}$, and $R$ is the gas constant, written as $R = c_p - c_v$. The properties $c_p$ and $c_v$ are the specific heat at constant pressure and volume, respectively. The modified viscous stress tensor may be written as
\begin{equation}
\tau_{ij}^{mod}=(\mu + \mu_{SGS}) \left(\frac{\partial \tilde{u}_i}{\partial x_j} + \frac{\partial \tilde{u}_j}{\partial x_i} \right) - \frac{2}{3} (\mu + \mu_{SGS}) \left(\frac{\partial \tilde{u}_k}{\partial x_k} \right) \delta_{ij} 
\end{equation}
where $\mu$ is the dynamic viscosity coefficient, calculated by Sutherland's Law, and $\mu_{SGS}$ is the SGS dynamic viscosity coefficient, which is provided by the subgrid-scale model. The strategy of modeling the subgrid-scale contribution as an additional dynamic viscosity coefficient is based on the Boussinesq hyphotesis. The modified heat flux vector, using the same modeling strategy, is given by
\begin{equation}
q_i^{mod}=-(k+k_{SGS})\frac{\partial \tilde{T}}{\partial x_i}
\end{equation}
where $k$ is the thermal conductivity coefficient of the fluid and $k_{SGS}$ is the SGS thermal conductivity coefficient given by
\begin{equation}
k_{SGS}=\frac{\mu_{SGS} c_p}{Pr_{SGS}}
\end{equation}
and $Pr_{SGS}$ is the SGS Prandtl number.
%
%
The work of \citet{Junior2018} compares the effects of different SGS models on the 
simulations of jet flows, and it observes a weak influence of the models. Therefore, 
the present work only applies the static Smagorinsky model \citep{Smagorinsky1963} in 
order to calculate the subgrid-scale contribution.

\subsection{Nodal Discontinuous Galerkin Method}
The nodal Discontinuous Galerkin method used in this work is obtained initially by dividing the physical domain $\Omega$ into multiple non-overlapping elements. Galerkin methods are capable of handling any element type, however, it is decided to restrict the shape of the elements to hexahedra. When working with hexahedral elements, the implementations is simpler and the computational efficiency is improved when compared to arbitrary shape elements.

The elements from the physical domain are mapped onto a reference unit cube elements $E=[-1,1]^3$. The equations, presented in (\ref{eq.1}) need also to be mapped to this new reference domain, leading to
\begin{equation}
J \frac{\partial \mathbf{\bar{Q}}}{\partial t} + \nabla_{\xi} \cdot \bar{\mathcal{F}} = 0,
\label{eq.2}
\end{equation}
where $\nabla_{\xi}$ is the divergence operator with respect to the reference element coordinates, $\mathbf{\xi}=(\xi^1,\xi^2,\xi^3)^T$, $J= \arrowvert \partial \mathbf{x} / \partial \mathbf{\xi} \arrowvert$ is the Jacobian of the coordinate transformation and $\bar{\mathcal{F}}$ is the contravariant flux vector.

The discontinuous Galerkin formulation is obtained multiplying (\ref{eq.2}) by the testfunction $\psi=\psi(\xi)$ and integrating over the reference element $E$
\begin{equation}
\int_E J \frac{\partial \mathbf{\bar{Q}}}{\partial t} \psi d \xi + \int_E \nabla_{\xi} \cdot \bar{\mathcal{F}} \psi d \xi = 0.
\label{eq.3}
\end{equation}
It is possible to obtain the weak form of the scheme by integrating by parts the second term in (\ref{eq.3})
\begin{equation}
\frac{\partial}{\partial t} \int_E J \mathbf{\bar{Q}} \psi d \xi + \int_{\partial E} (\bar{\mathcal{F}} \cdot \vec{N})^* \psi dS - \int_E \bar{\mathcal{F}} \cdot (\nabla_{\xi} \psi ) d \xi = 0,
\label{eq.4}
\end{equation}
where $\vec{N}$ is the unit normal vector of the reference element faces. Because the discontinuous Galerkin scheme allows discontinuities in the interfaces, the surface integral above is ill-defined. In this case, a numerical flux, $\bar{\mathcal{F}}^*$, is defined, and a Riemann solver is used to compute the value of this flux based on the discontinuous solutions given by the elements sharing the interface.

For the nodal form of the discontinuous Galerkin formulation, the solution in each element is approximated by a polynomial interpolation of the form
\begin{equation}
\mathbf{\bar{Q}}(\xi) \approx \sum_{p,q,r=0}^N \mathbf{\bar{Q}}_h(\xi_p^1,\xi_q^2,\xi_r^3,t)\phi_{pqr}(\xi),
\end{equation}
where $\mathbf{\bar{Q}}_h(\xi_p^1,\xi_q^2,\xi_r^3,t)$ is the value of the vector of conserved variables at each interpolation node in the reference element and $\phi_{pqr}(\xi)$ is the interpolating polynomial. For hexahedral elements, the interpolating polynomial is a tensor product basis with degree N in each space direction
\begin{equation}
\phi_{pqr}(\xi)=l_p(\xi^1)l_q(\xi^2)l_r(\xi^3), \hspace{10pt} l_p(\xi^1)= \prod_{\substack{i=0 \\ i \ne p}}^{N_p} \frac{\xi^1-\xi_i^1}{\xi_p^1-\xi_i^1}.
\end{equation}
The definitions presented are applicable to other two directions.

The numerical scheme used in the simulations have a split formulation, as presented by \citet{Pirozzoli2011}, with the discrete form given by \citet{Gassner2016}, to enhance the stability of the simulation. The split formulation is employed to Euler fluxes only. The solution and the fluxes are interpolated and integrated at the nodes of a Gauss-Lobatto Legende quadrature, which presents the summation-by parts property, that is necessary to employ the split formulation. 

The Riemann solver used in the simulations is a Roe scheme with entropy fix \citep{Harten1983} to ensure that second law of thermodynamics is respected, even with the split formulation. For the viscous flux, since the discontinuous Galerkin scheme are not suitable for discretizing the high order derivative operator, the lifting scheme of \citet{BassiRebay1997} is used. The time marching method chosen is the five-stage, fourth-order explicit Runge-Kutta scheme of \citet{CarpenterKennedy1994}. The shock waves that appear in the simulation are stabilized by using the sub-cell shock capturing method of \citet{Sonntag2017}.

\subsection{Second-Order Finite Difference Method}

The code JAZzY \citep{Junior2018} solves the non-dimensional filtered Navier-Stokes 
equations using a structured finite difference approach for a general curvilinear 
coordinate system. It calculates the numerical fluxes using a second-order centered 
scheme with explicit addition of artificial dissipation, which is obtained from the 
anisotropic scalar model of \citet{TurkelVatsa1994}. The time integration is done by 
an explicit second-order five-stage Runge-Kutta scheme \citep{Jameson1986}. Even 
though the simulation is performed for a supersonic problem, no additional 
shock-capturing model is required, because the \citet{TurkelVatsa1994} artificial 
dissipation model can provide additional artificial dissipation near discontinuities

\section{EXPERIMENTAL CONFIGURATION}

Most of recent jet experiments are concerned with representing the sound emitted 
from modern airplanes in takeoff configuration, in which the jet flow is in high 
subsonic regime. However, our focus is to represent supersonic jet flows, that can 
be found in large launch vehicles. The experimental work of \citep{BridgesWernet2008} 
provided the physics characteristics desired for the simulation in which 
a group of nozzle configurations is analyzed to obtain different types of jet flows.

The present work studies the fully expanded free jet flow configuration with a Mach 
number of $1.4$. This choice is motivated by the lack of wall geometries and the 
absence of strong shock waves in the configuration. Such flow characteristics can 
contribute to reducing the calculation costs. The experimental apparatus for this 
configuration is composed of a convergent-divergent nozzle designed with the method 
of characteristics to obtain this flow configuration \citet{BridgesWernet2008}. The 
nozzle exit diameter is $50.8$ mm. The Reynolds number based on the nozzle exit 
diameter reaches values of approximately $1.65 \times 10^{6}$, which is large compared 
to most other jet experiments and LES available in the literature.

The data acquisition in the tests applies Time-Resolved Particle Image Velocimetry 
(TRPIV), and it is operated primarily with a 10 kHz sample rate. For spectra analysis, 
in some smaller fields, the acquisition is made with 25 kHz. The experiment uses two 
sets of cameras, one positioned to better capturing the flow along the nozzle 
centerline and the other to capture the flow of the mixing layer along the nozzle 
lipline.

\section{NUMERICAL SETUP}

\subsection{Geometry Characteristics}

In the present work, the authors decided to utilize the geometry that provided 
the best results in the work of \citet{Junior2018}. Such a configuration has a 
divergent shape with an axis length of $40 D$. Close to the jet inlet, one can 
find a smaller diameter of $16 D$ and, on the opposite side, it presents a 
diameter of $25 D$. The code JAZzY is non-dimensional and the reference length 
is the exit nozzle diameter $D$. The simulations using FLEXI are dimensional. 
Therefore, the geometry is designed with the same scale as the experiment, so 
the nozzle exit diameter in the geometry has $D = 50.8$ mm. A 3-D view of the 
geometry domain is presented in Fig.\ \ref{fig.1a}.

\subsection{Mesh Configuration}

\begin{figure}[htb!]
\centering
\subfloat[]{
	\includegraphics[width=0.4\linewidth]{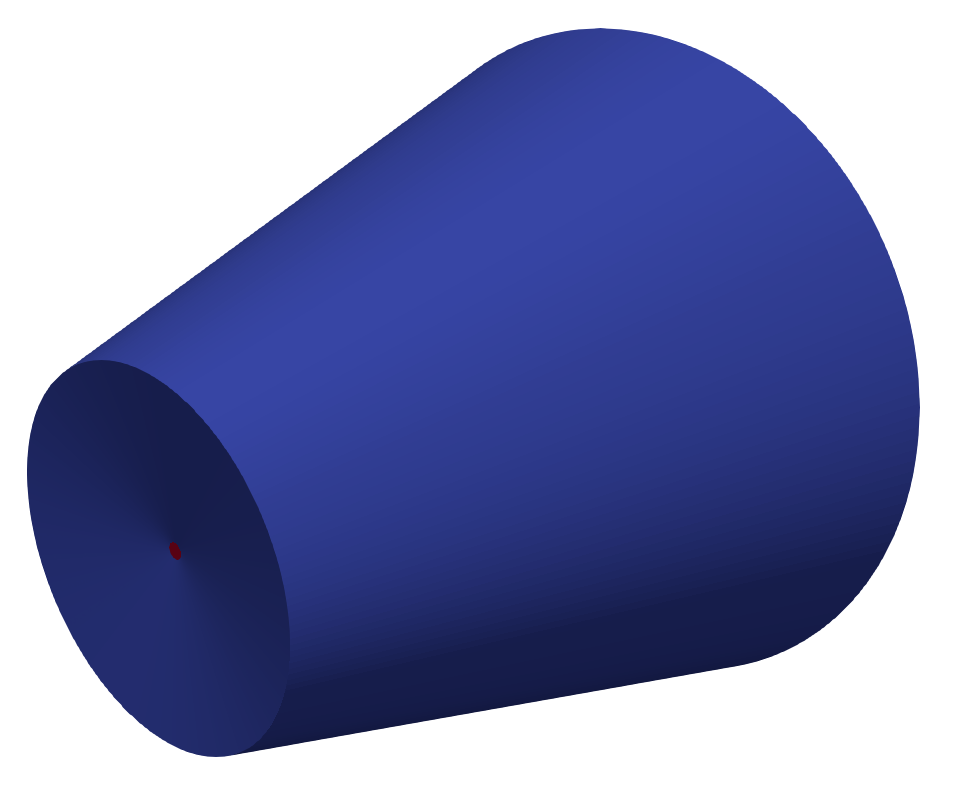}
	\label{fig.1a}
	}
\subfloat[]{
	\includegraphics[width=0.55\linewidth]{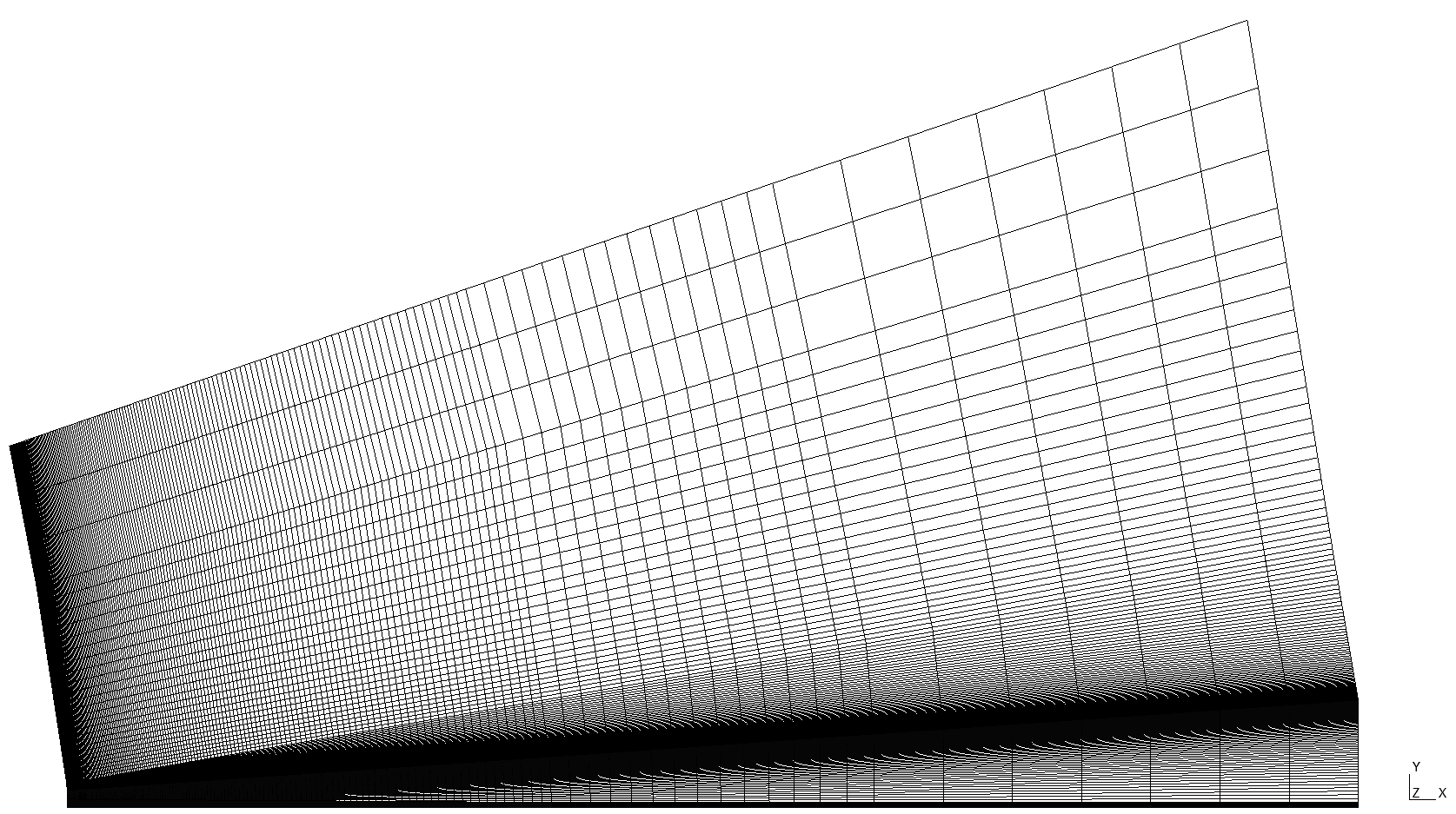}
	\label{fig.1b}
	}
\caption{(a): 3D view of geometry domain coloured by boundary condition. Red for jet inlet boundary condition and blue for the domain boundary condition. (b): Half cutplane of Flexi-2 mesh.}
\end{figure}

The JAZzY code works with a geometry that is obtained from the 
rotation of a 2-D mesh that creates prismatic elements connecting 
the centerline. FLEXI code is restricted to hexahedral elements. Hence, in 
order to create two similar grids, the procedure of rotating a 2-D mesh was 
done differently. Instead of having a 2-D mesh starting in the centerline, 
it started with a diameter of $0.30D$ and the interior of the mesh is filled 
with a rectangular core connected to the rotated mesh. This strategy allows 
obtaining the same mesh topology for diameters larger than $0.30D$. A cut plane 
of the mesh used in the FLEXI simulations is presented in Fig.\ \ref{fig.1b}.

Due to the use of different numerical algorithms and orders of accuracy, it is
necessary to establish comparison criteria for the simulations. For the JAZzY 
code, the number of degrees of freedom in each cell is one, because it only has 
one value of the properties in each cell. For the FLEXI code, the number of 
degrees of freedom in each cell depends on the degree of chosen interpolating 
polynomial. For a second-order simulation, we have a first-order polynomial, 
that requires two internal nodes in each direction, and consequently, the 
degree of freedom in each element is $8$. For a third-order simulation, $3$ 
internal nodes are required in each simulation, leading to $27$ degrees of 
freedom in each element.

Considering the degrees of freedom as the comparison criteria and the 
reference grid provided in \citet{Junior2018}, with approximately $49 \times 
10^{6}$ points, the two meshes created for the FLEXI code present approximately 
$6 \times 10^{6}$ and $8 \times 10^{6}$ elements. The adjustment of mesh sizes 
is obtained uniformly through the whole domain to guarantee consistency between 
the simulations. Table \ref{tab.1} presents a summary of the mesh refinements 
used in the current work. The present work uses GMSH \citep{Geuzaine2009} to 
generate the computational grids.
\begin{table}[htb!]
\centering
\caption{Summary of the meshes utilized in the simulations.}
\begin{tabular}{ c | c | c | c | c } \hline
Meshes & Order of Accuracy & DOF/cell & Cells ($10^{6}$) & Total \# of DOF ($10^{6}$) \\ \hline
JAZzY & 2nd order & 1 & $49.1$ & $49.1$ \\
Flexi-2 & 2nd order & 8 & $6.2$ & $49.6$ \\
Flexi-3  & 3rd order & 27 & $1.8$ & $48.6$ \\ \hline
\end{tabular}
\label{tab.1}
\end{table}

\subsection{Boundary Conditions}

The present calculations apply the same boundary conditions of \citet{Junior2018} 
for the sake of consistency. For the jet inlet boundary condition, the red region in 
Fig.\ \ref{fig.1a}, all properties are prescribed and a top hat velocity profile 
is imposed. In the JAZzY simulation, the inflow boundary conditions imposed a 
non-dimensional jet velocity of $1.4$, with non-dimensional pressure and temperature 
of $1.0$, and a Reynolds number of $1.57 \times 10^{6}$. The inflow boundary 
condition of FLEXI simulations imposes a jet velocity of $476.37$ m/s, with a 
pressure of $101,325$ Pa and a density of $1.225$ kg/m$^3$, giving a Reynolds 
number based on the jet diameter of $1.66 \times 10^{6}$.

For the domain boundary condition, the blue region in Fig.\ \ref{fig.1a}, Riemann 
invariant-type boundary conditions are used to characterize the farfield in the 
JAZzY code, while Dirichlet boundary conditions are used in FLEXI for the farfield 
boundary. A sponge zone \citep{Flad2014} is used in the simulations with FLEXI as an 
attempt to maintain the non-reflective properties of the Riemann invariants. This 
sponge region is implemented circumferentially, with a small damping value starting at 
approximately $14D$ and increased at longer distances from the centerline.


\subsection{Calculation of Statistical Properties}

For all the simulations, the initial condition is a stagnated flow all over the 
domain. Then, the jet is fully developed with approximated $3$ flow-through times 
(FTT). One FTT is defined as the time taken by a particle with the jet velocity 
to cross the domain. For another $3$ FTTs, the developed jet can reach the statistically 
steady condition. The statistics presented in this work are achieved from the following 
$3$ FTTs. The complete simulation takes $9$ FTTs. The JAZzY simulations run $14.2$ FTTs 
before starting to collect statistical data for further $3.3$ FTTs.

In the FLEXI simulations, the data are stored with a frequency of 100 kHz, which is at least 
4 times larger than the acquisition frequency from the experiments. From the work of 
\citet{Junior2018}, the data from the simulation using JAZzY are stored with a 
frequency of 280 kHz.

\section{RESULTS}

Two simulations are performed using the FLEXI code in the present work. The first 
calculation, Flexi S1, uses second-order accuracy and applies the Flexi-2 mesh, as 
described in Tab.\ \ref{tab.1}. The second simulation, Flexi S2, runs with third-order 
spatial discretization and the Flexi-3 mesh, which has its properties also indicated in 
Tab.\ \ref{tab.1}. The results of the simulations are compared with numerical \citep{Junior2018} 
and experimental data \citep{BridgesWernet2008}.

The simulations were performed using 3600 computational cores from the Jean-Zay supercomputer 
\citep{JeanZay2021}. The high scalability performance of the FLEXI code is discussed in the work of 
\citet{Hindenlang2012} and \citet{Krais2021}. One interesting metric to compare the performance 
of the two numerical spatial discretization methods is the performance index (PID). 
The performance index is the time 
needed by the simulation to advance a single DOF one stage of the Runge-Kutta time
integration scheme. It can be computed by
\begin{equation}
PID = \frac{wall \hspace{2pt} clock \hspace{2pt} time \cdot n_{cores}}{n_{DOF} \cdot n_{time \hspace{1pt} steps} \cdot n_{RK-stages}}.
\end{equation}
The simulation using the FLEXI code, running with 2nd-order accuracy in the Flexi-2 mesh, presented a PID 
of 7 $\mu$s/DOF. Similarly, the code running with the 3rd-order accurate option, using the Flexi-3 mesh, 
presented a PID of 13 $\mu$s/DOF. 
The JAZzY code also presents high scalability and detailed studies of the parallel performance of the 
code are presented in the work of \citet{Junior2019} and \citet{Junior2020}.

\begin{figure}[htb!]
\centering
\subfloat[Centerline]{
	\includegraphics[width=0.49\linewidth]{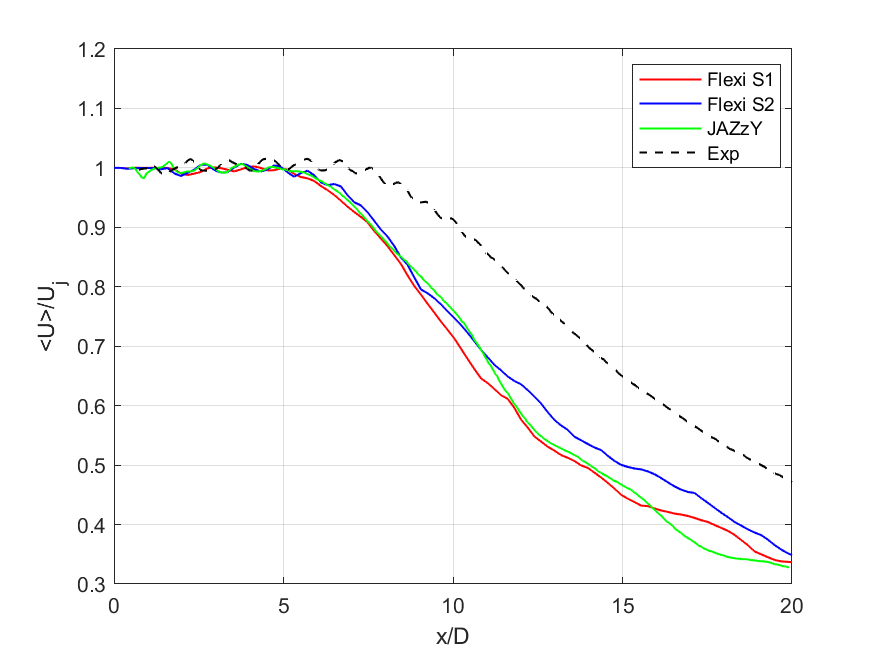}
	\label{fig.2a}	
	}%
\subfloat[Centerline]{
	\includegraphics[width=0.49\linewidth]{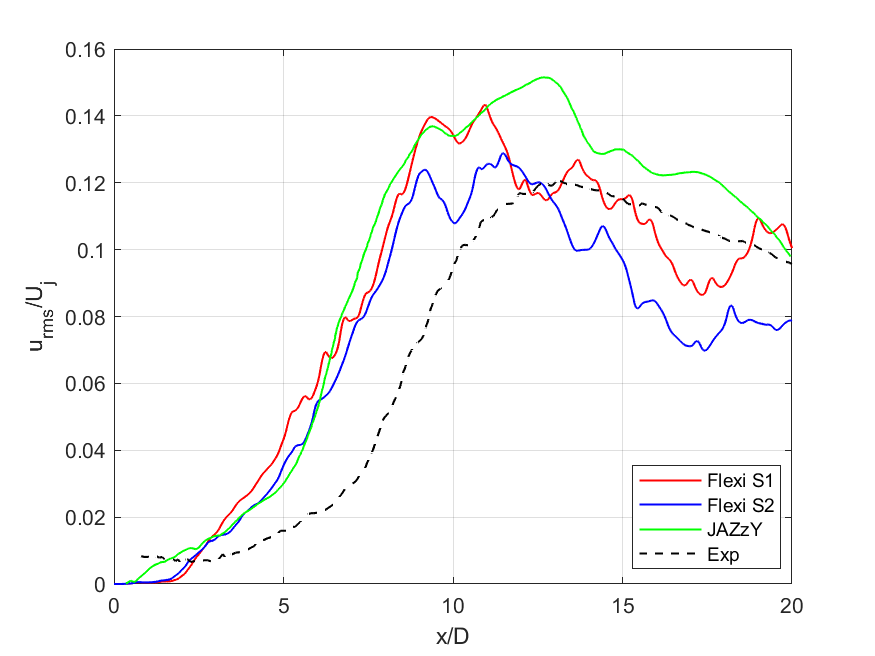}
	\label{fig.2b}	
	}
\newline
\subfloat[Lipline]{
	\includegraphics[width=0.49\linewidth]{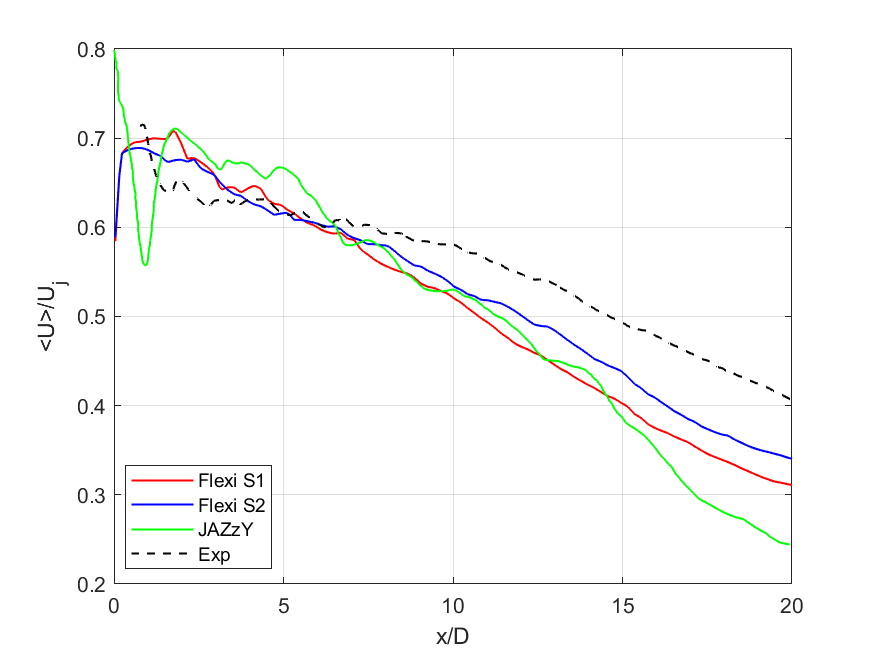}
	\label{fig.2c}
	}%
\subfloat[Lipline]{
	\includegraphics[width=0.49\linewidth]{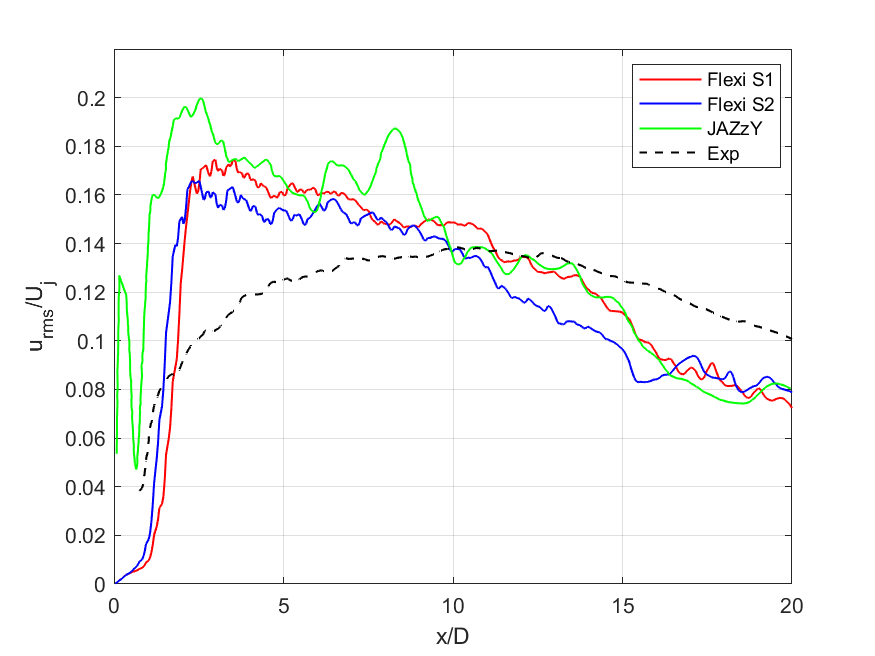}
	\label{fig.2d}	
	}
\caption{Results of mean longitudinal velocity component distribution (left) and rms of longitudinal velocity fluctuation (right) in the jet centerline $y/D=0$ (top) and lipline $y/D=0.5$ (bottom).}
\label{fig.2}
\end{figure}

The first results investigated are the distribution of mean longitudinal 
velocity component (<$U$>) and the distribution of root mean square (rms) of 
longitudinal velocity component fluctuation ($u_{rms}$) along the jet centerline 
($y/D=0$) and the jet lipline ($y/D=0.5)$. The velocity variables are 
non-dimensionalized by dividing by jet velocity $U_j$. The results are presented 
in Fig.\ \ref{fig.2}. It is possible to observe from these results that velocity 
profiles calculated using the FLEXI numerical tool are very close to the one from 
JAZzY computation. Such a remark indicates little influence of the numerical 
method and order of accuracy for the conditions of such flow configuration.

The mean velocity distribution in the jet centerline, provided by the FLEXI 
simulations, Fig.\ \ref{fig.2a}, presents a similar behavior of JAZzY calculations 
with a shorter potential core than the one observed in the experiments. After the 
velocity starts decreasing, the slope of the profile achieved by the simulations 
and the experimental measures is very similar. The shorter potential core may be 
correlated with the earlier increase in the velocity fluctuation, as presented in 
Fig.\ \ref{fig.2b}, that all three numerical studies produced. In all simulations, 
it is possible to observe that the peaks in the calculated longitudinal velocity 
component fluctuations occur upstream of the one observed in the experimental data. 
Furthermore, such peaks in the computational results are considerably more pronounced 
than the ones observed in the experiments. The simulation with third-order of accuracy 
generates the smallest peaks among all calculations. However, it produces the largest 
double-peaked amplitude, which is a behavior reproduced by all numerical simulations 
that do not exist in the experiments.

The mean longitudinal velocity component distribution along the lipline, generated 
by all three calculations, Fig.\ \ref{fig.2c}, once more show different results, 
presenting an earlier decrease in the velocity. Figure \ref{fig.2d} presents similar 
behavior for the rms values of velocity fluctuations in the lipline calculated using JAZzY 
and FLEXI, which present a sudden peak closer to the nozzle exit station that slowly 
decreases, while the experiment shows a slow increase of fluctuation velocity and 
the occurrence of the peak fluctuation farther from the nozzle exit station. The 
magnitude of the peak presented in the numerical simulations is significantly higher 
than the one presented in the experiments.

These results indicate an excessive turbulent intensity close to the nozzle outlet. 
Such a remark is possibly due to the choice of the inviscid top-hat velocity profile 
imposed at the inlet boundary condition. Other jet studies \citep{BogeyBailly2010} 
indicate significant differences in the results due to the inflow condition. They 
run simulations changing the boundary layer thickness for initially laminar jets and 
notice that smaller boundary layer thickness generates sooner peaks of velocity 
fluctuation, which is probably the behavior observed in the simulations of this 
work.

\begin{figure}[htb!]
\centering
\subfloat[$x/D=2.5$]{
	\includegraphics[trim = 30mm 0mm 40mm 0mm, clip, width=0.207\linewidth]{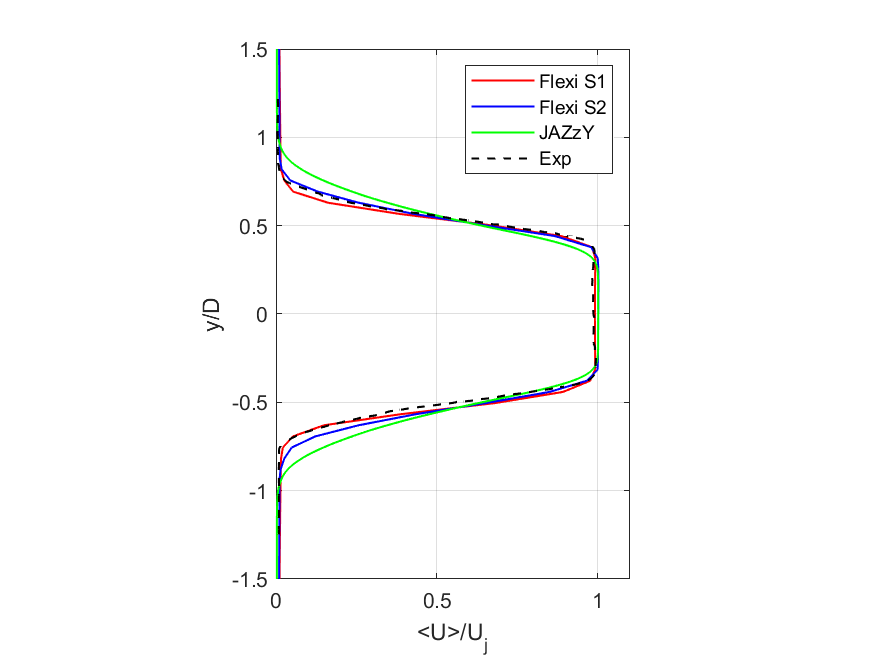}
	\label{fig.3a}	
	}
\subfloat[$x/D=5$]{
	\includegraphics[trim = 30mm 0mm 40mm 0mm, clip, width=0.207\linewidth]{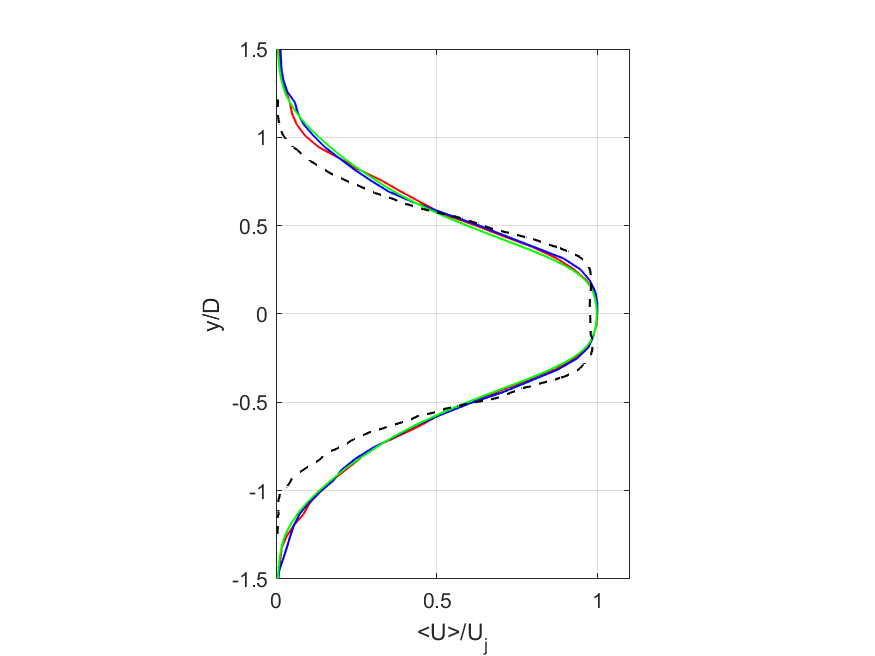}
	\label{fig.3b}	
	}
\subfloat[$x/D=10$]{
	\includegraphics[trim = 30mm 0mm 40mm 0mm, clip, width=0.207\linewidth]{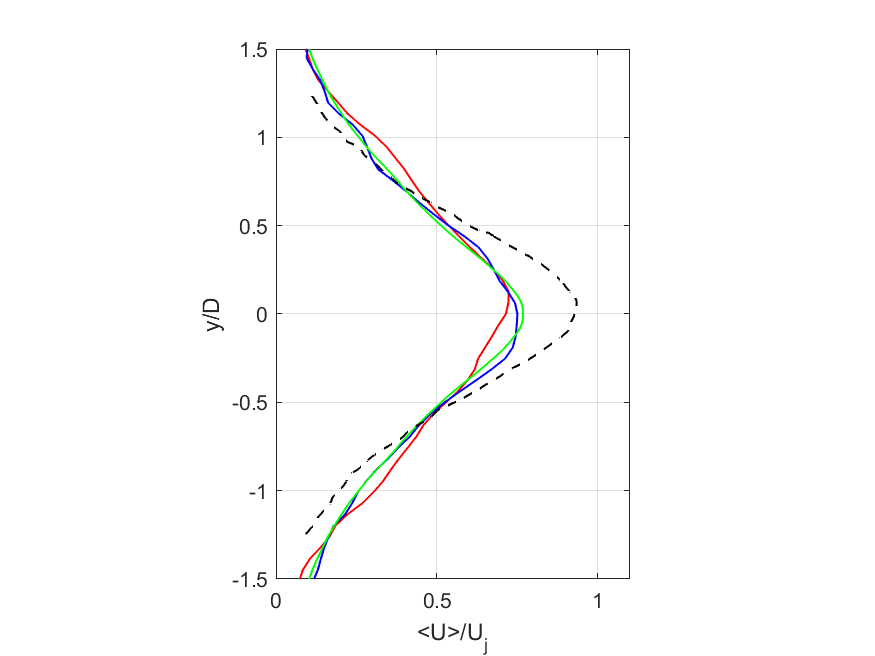}
	\label{fig.3c}
	}
\subfloat[$x/D=15$]{
	\includegraphics[trim = 30mm 0mm 40mm 0mm, clip, width=0.207\linewidth]{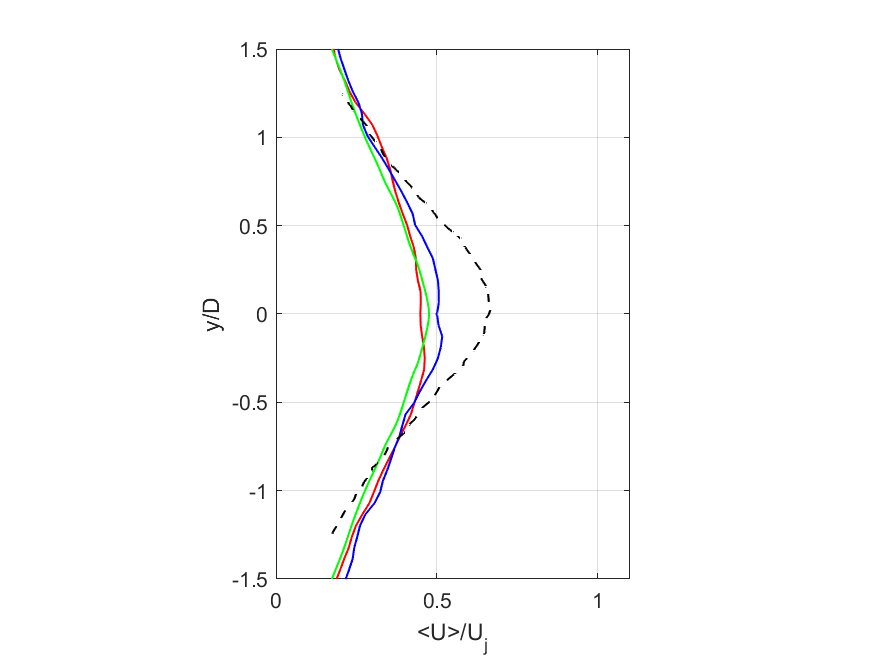}
	\label{fig.3d}	
	}
\newline
\subfloat[$x/D=2.5$]{
	\includegraphics[trim = 30mm 0mm 40mm 0mm, clip, width=0.207\linewidth]{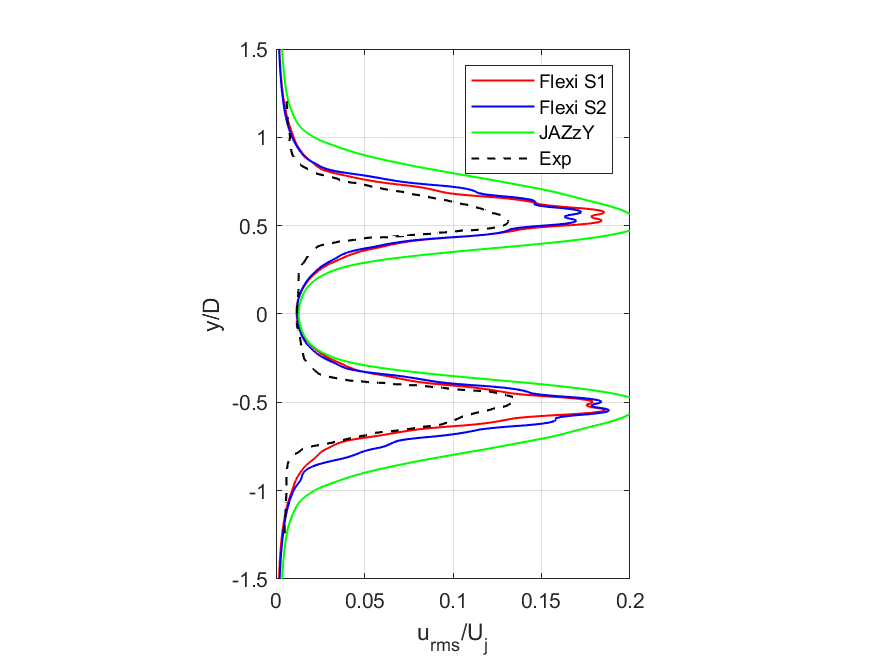}
	\label{fig.3e}	
	}
\subfloat[$x/D=5$]{
	\includegraphics[trim = 30mm 0mm 40mm 0mm, clip, width=0.207\linewidth]{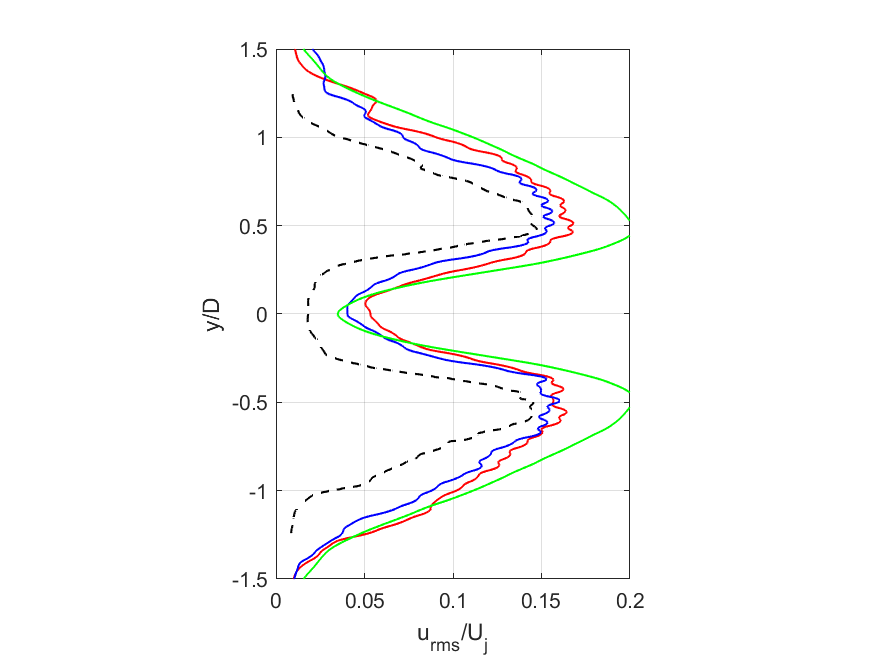}
	\label{fig.3f}	
	}
\subfloat[$x/D=10$]{
	\includegraphics[trim = 30mm 0mm 40mm 0mm, clip, width=0.207\linewidth]{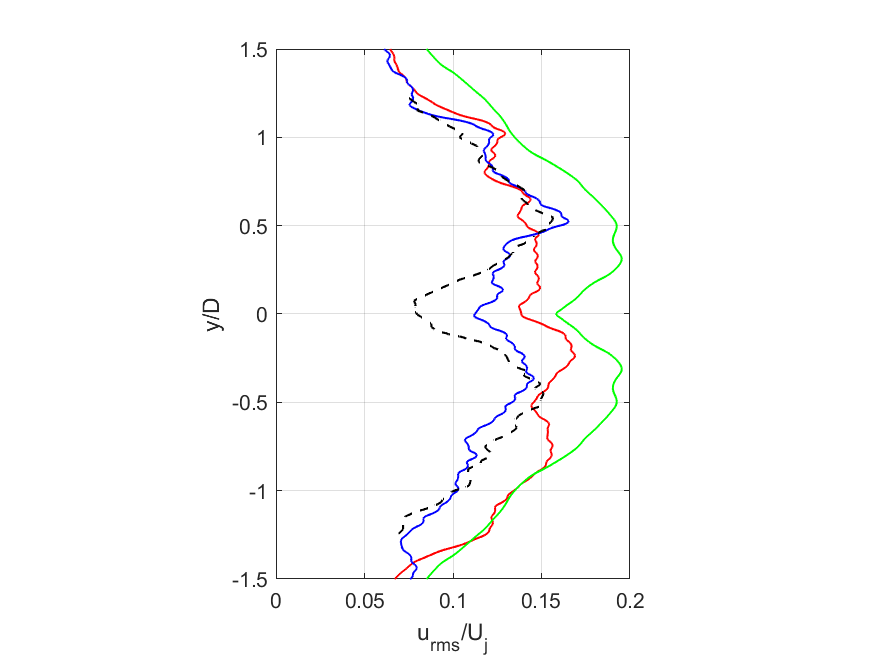}
	\label{fig.3g}
	}
\subfloat[$x/D=15$]{
	\includegraphics[trim = 30mm 0mm 40mm 0mm, clip, width=0.207\linewidth]{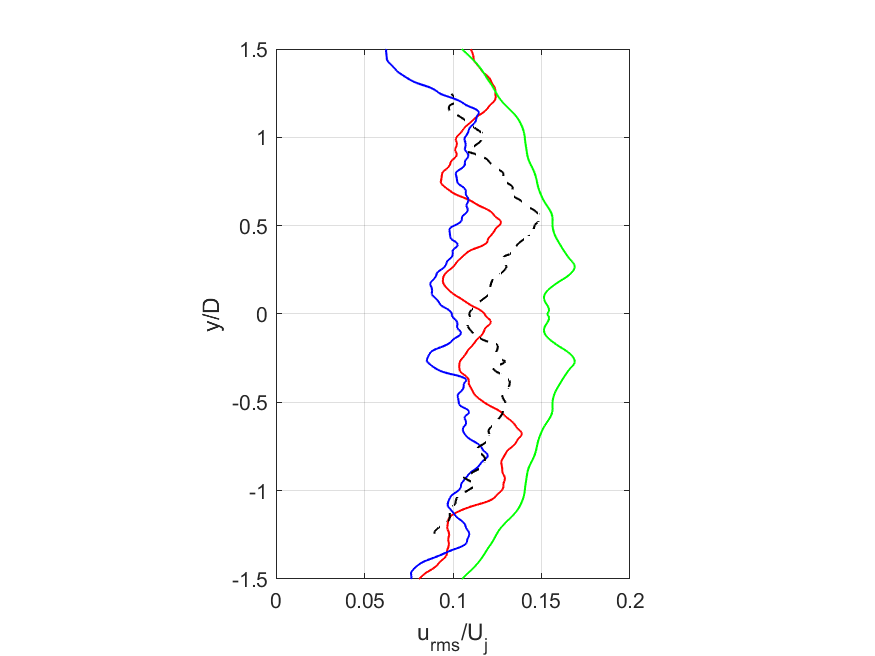}
	\label{fig.3h}	
	}
\newline
\subfloat[$x/D=2.5$]{
	\includegraphics[trim = 30mm 0mm 40mm 0mm, clip, width=0.207\linewidth]{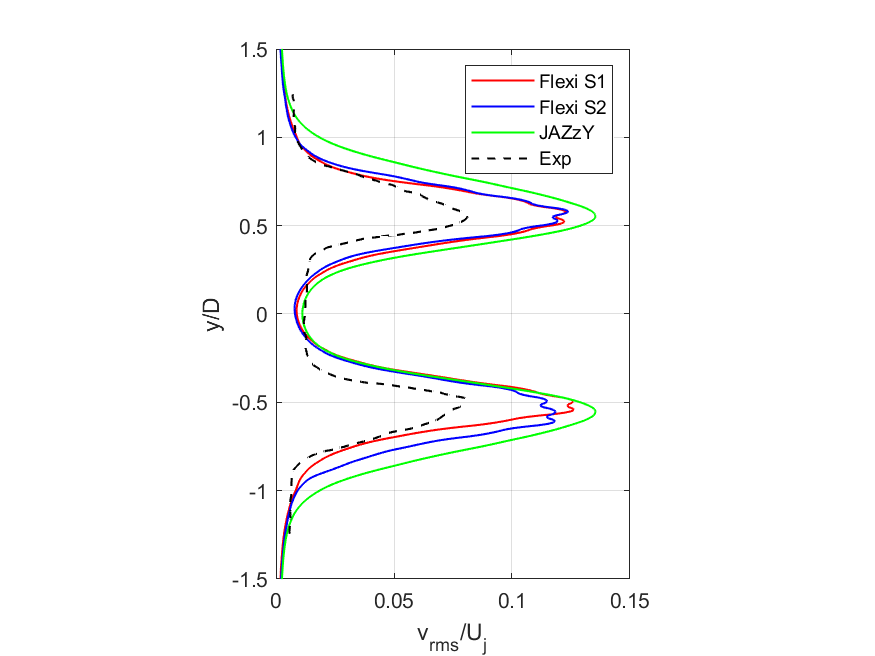}
	\label{fig.3i}	
	}
\subfloat[$x/D=5$]{
	\includegraphics[trim = 30mm 0mm 40mm 0mm, clip, width=0.207\linewidth]{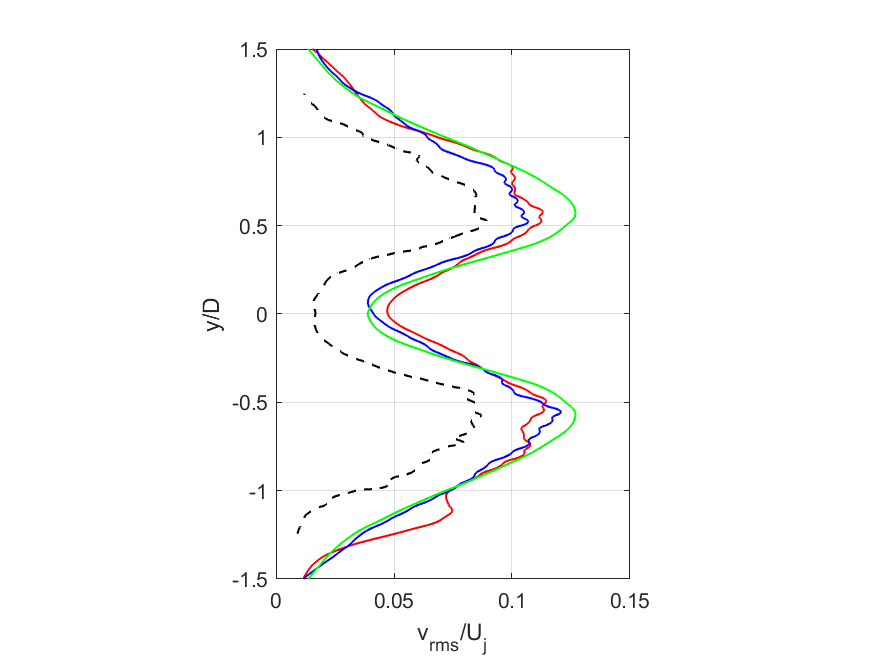}
	\label{fig.3j}	
	}
\subfloat[$x/D=10$]{
	\includegraphics[trim = 30mm 0mm 40mm 0mm, clip, width=0.207\linewidth]{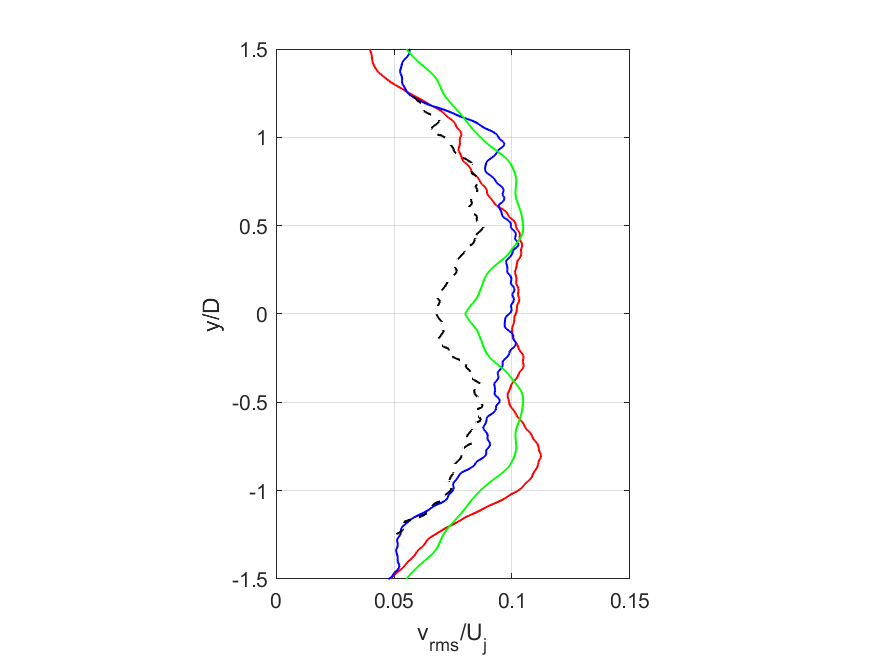}
	\label{fig.3k}
	}
\subfloat[$x/D=15$]{
	\includegraphics[trim = 30mm 0mm 40mm 0mm, clip, width=0.207\linewidth]{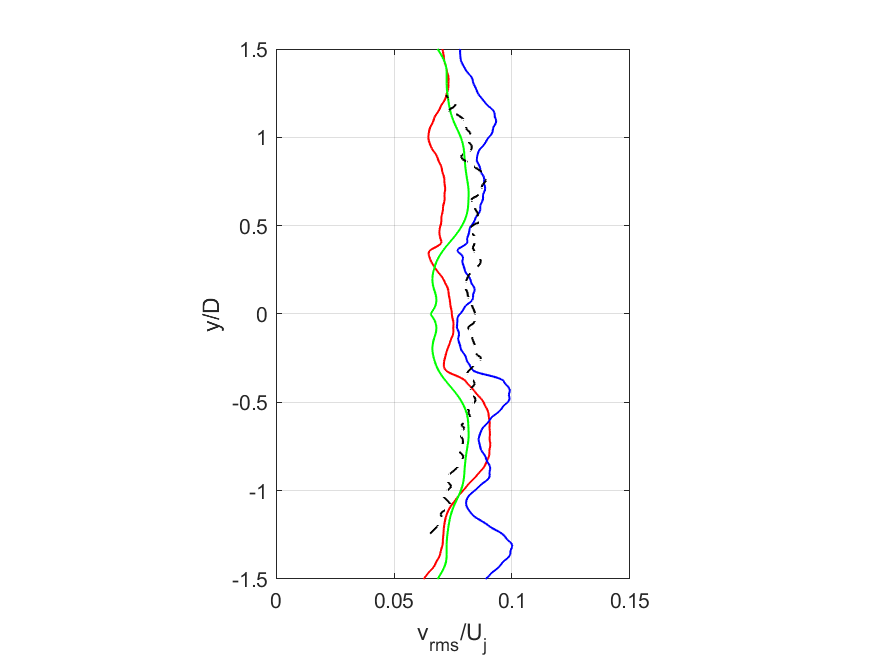}
	\label{fig.3l}	
	}
\newline
\subfloat[$x/D=2.5$]{
	\includegraphics[trim = 30mm 0mm 40mm 0mm, clip, width=0.207\linewidth]{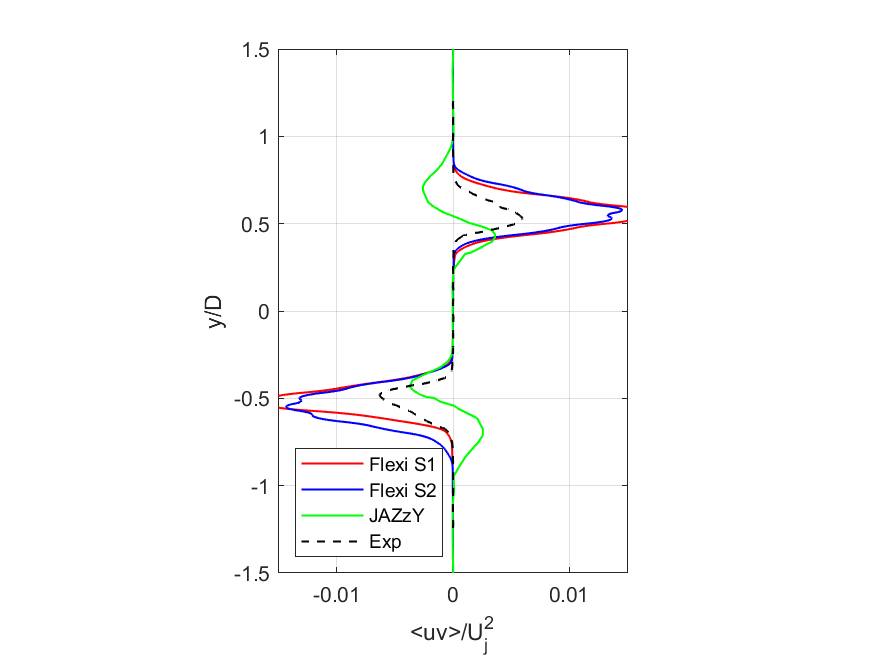}
	\label{fig.3m}	
	}
\subfloat[$x/D=5$]{
	\includegraphics[trim = 30mm 0mm 40mm 0mm, clip, width=0.207\linewidth]{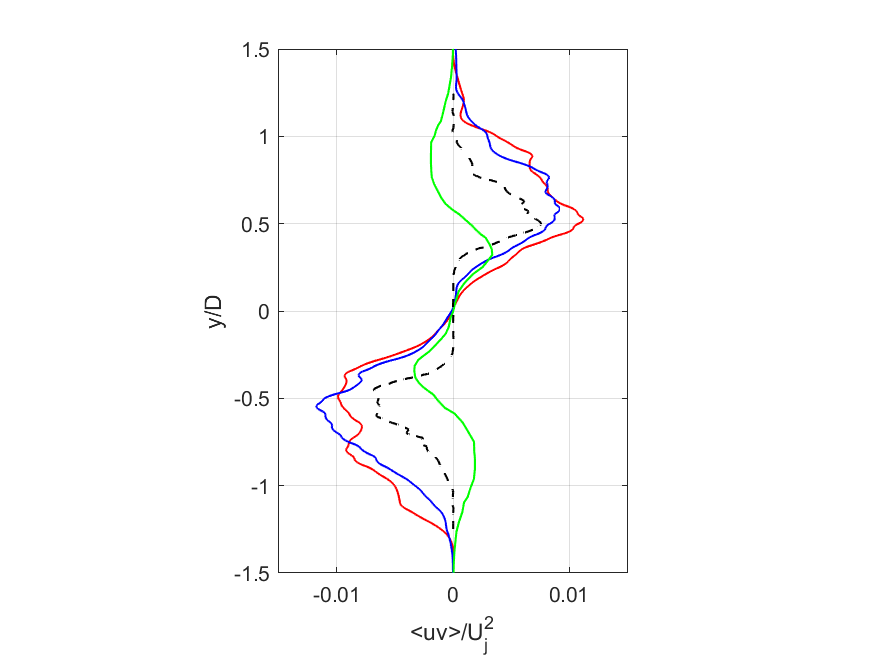}
	\label{fig.3n}	
	}
\subfloat[$x/D=10$]{
	\includegraphics[trim = 30mm 0mm 40mm 0mm, clip, width=0.207\linewidth]{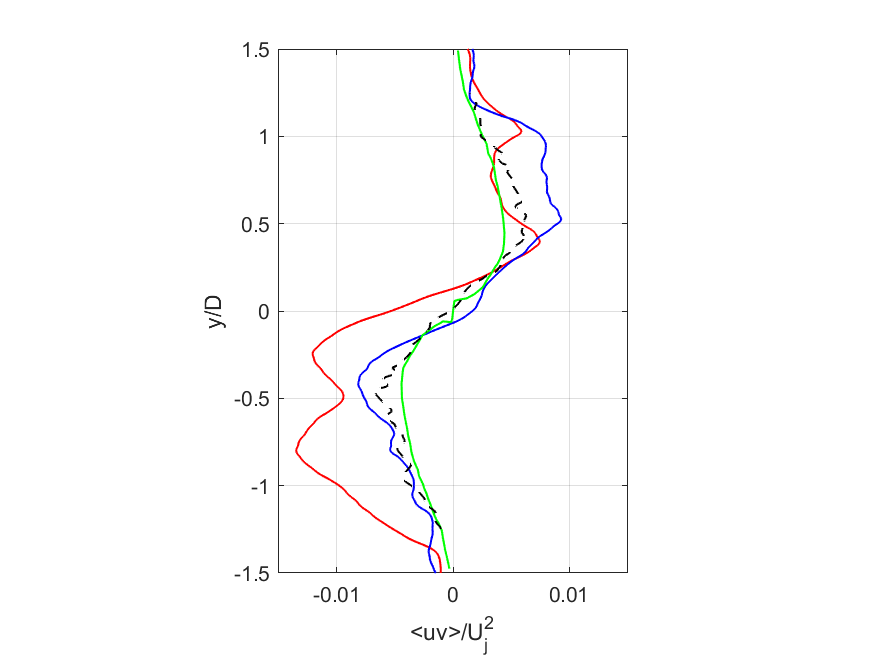}
	\label{fig.3o}
	}
\subfloat[$x/D=15$]{
	\includegraphics[trim = 30mm 0mm 40mm 0mm, clip, width=0.207\linewidth]{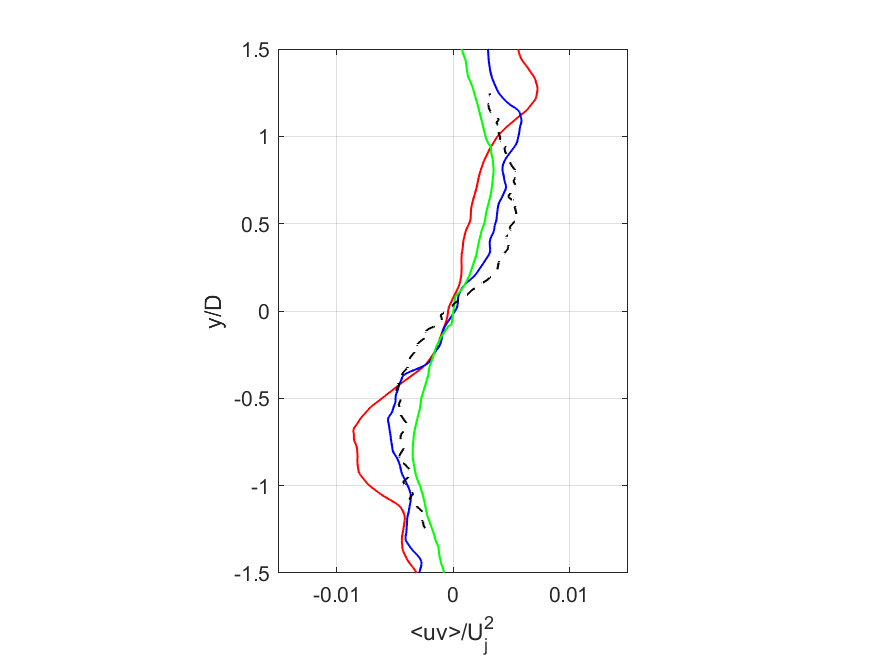}
	\label{fig.3p}	
	}
\caption{Results of mean longitudinal velocity component, rms of longitudinal velocity fluctuation, rms of transversal velocity fluctuation and mean Reynolds shear-stress tensor component (from top to bottom) are presented in four streamwise positions $x/D=2.5$, $x/D=5$, $x/D=10$ and $x/D=15$ (from left to right).}
\label{fig.3}
\end{figure}

Figure \ref{fig.3} presents the mean and the rms fluctuation of the longitudinal 
velocity profiles in four planes streamwise of nozzle flow at positions $x/D=2.5$, 
$x/D=5$, $x/D=10$ and $x/D=15$. Additional results of rms of transversal velocity 
fluctuation ($v_{rms}$) distribution and mean Reynolds shear-stress tensor component 
(<$uv$>) distribution are presented in the same figure. The results of transversal 
velocity fluctuation and Reynolds shear-stress tensor components are also 
non-dimensionalized by jet velocity.

The mean results of longitudinal velocity distribution are analyzed in the four 
planes, Figs.\ \ref{fig.3a}-\ref{fig.3d}. In the first plane, Fig.\ \ref{fig.3a}, 
the results indicate a good agreement between the simulations and experimental 
data, which indicates little influence of the inlet condition at the beginning 
of the jet. Downstream, in Fig.\ \ref{fig.3b}, it is observed a larger spreading 
of velocity when comparing to the experimental reference, with peaks that are 
closer to the measured data when compared to the JAZzY profile. The results 
presented in the last two positions, Figs.\ \ref{fig.3c} and \ref{fig.3d}, show 
similar behavior, a large spreading of the velocity with a smaller peak than the 
one presented by experimental data.

Figures \ref{fig.3e} to \ref{fig.3h} illustrate results from 
rms of longitudinal velocity fluctuation. In the first position, 
Fig.\ \ref{fig.3e}, it is observed the largest peak of fluctuation 
close to the lipline ($y/D=0.5$) for all simulations. JAZzY 
simulation results indicate the highest peaks among all 
computational data. All calculations present values of $u_{rms}/U_j$ 
very close to experimental at the centerline ($y/D=0$). These 
results show that, very close to the nozzle exit, even with almost 
twice of velocity fluctuation, the mean velocity results are kept 
the same.
The results of the second plane position, Fig.\ \ref{fig.3f}, show 
that the largest velocity fluctuations produced at the lipline have 
already been transported to the centerline producing higher values 
of fluctuation than the ones observed in the experiments. Similar 
to what is observed in Fig.\ \ref{fig.3e}, the results of JAZzY 
simulations still present higher levels of fluctuation than the 
ones obtained by FLEXI simulations. In the third plane, Fig.\ 
\ref{fig.3g}, while the experimental data indicates that the 
fluctuation at the centerline has not yet achieved the levels of 
the lipline, in all the simulations the velocity fluctuation of the 
centerline presents similar values to the values observed at the 
lipline. The velocity fluctuations from the JAZzY simulation are 
still higher than the values obtained from FLEXI simulations, 
which are similar to the velocity fluctuation of the experiment. 
Figure \ref{fig.3h} shows similar velocity fluctuations 
distributions transversally with the JAZzY simulations presenting 
the highest values.

The results from rms of transversal velocity fluctuation, presented 
in Figs.\ \ref{fig.3i} to \ref{fig.3l}, show similar distribution 
observed in the results from rms of longitudinal velocity distribution, 
Figs.\ \ref{fig.3e} to \ref{fig.3h}. The only difference is the fact 
that JAZzY and FLEXI results are ver similar for $v_{rms}/U_j$ profiles.

The mean of the Reynolds shear-stress tensor component is illustrated
in Figs.\ \ref{fig.3m} to \ref{fig.3p}. One can notice an important
mismatch of JAZzY results when comparing the FLEXI and experimental
profiles. Even in the first plane, Fig.\ \ref{fig.3m}, the values of 
the Reynolds shear-stress tensor components obtained from the JAZzY 
simulation are much smaller than the results from FLEXI simulations. 
Moreover, the shape of the tensor distribution is also different 
presenting peaks with inverted signs while the FLEXI simulations 
produce one peak with the same direction of the experimental data. 
The advance to farther planes from the inlet jet show the decrease 
in the values of the Reynolds shear-stress tensor component for the 
results of FLEXI simulations, reaching values close to experimental 
data, while the results from JAZzY simulation show a reduction of 
one of the peeks and a movement of the peek to the centerline.

The mean velocity results show almost no difference between
simulations performed in the present work and the 
reference calculations. The main differences are observed 
in the velocity fluctuations and the Reynolds shear-stress 
tensor component. The high levels of fluctuation produced 
in the JAZzY simulations may be related to the few artificial 
dissipation produced by the numerical method. The FLEXI 
simulations produced few levels of the fluctuations, which 
is similar to the levels produced in the experiment and the 
third order of accuracy simulation is the responsible for the 
few levels of velocity fluctuation. Even though, the third 
order accuracy is the most expensive simulation from all 
performed, it produced slightly better results for the same 
degree of freedom.

\FloatBarrier

\section{CONCLUSIONS}

In this work, the authors investigated different numerical 
methods to simulate supersonic jet flows. The authors 
performed an analysis of previous calculations using JAZzY 
code, which uses finite difference method with second-order 
accuracy, and observed higher levels of velocity fluctuation 
in the lipline and the centerline, which caused a shorter 
potential core of the jet. Another framework, FLEXI, based on 
discontinuous Galerkin formulation on unstructured meshes, is 
used in this work to simulate the same problem with 
second-order accuracy and also third-order accuracy. The 
results obtained from FLEXI simulations are very similar to 
the results from JAZzY simulations.

The results of all simulations produced high peaks of velocity 
fluctuation close to the nozzle outlet. This high level of 
velocity fluctuation is transported to the jet centerline 
sooner than obtained in the experiments and consequently 
increased the levels of velocity fluctuation along the 
centerline. This phenomenon can be responsible for the short 
potential core of the jets obtained in the simulations. This 
result is also obtained in other works in the literature that 
investigated different velocity profiles imposed in the 
inflow condition for the jet.

Little influence is observed with different numerical methods 
for a fixed number of degree of freedom. The results of the 
simulation with third-order accuracy are the closest to 
experimental data, indicating that it is a strong candidate 
to be used for other configurations. Considering the results 
obtained, with almost no influence of different numerical 
methods and order of accuracy, the continuity of the work 
should be directed to the imposition of the nozzle exit 
boundary condition.

\section{ACKNOWLEDGEMENTS}

The authors acknowledge the support for the present research provided by Conselho Nacional de Desenvolvimento Cient\'{\i}fico e Tecnol\'{o}gico, CNPq, under the Research Grant No.\ 309985/2013-7\@. The work is also supported by the computational resources from the Center for Mathematical Sciences Applied to Industry, CeMEAI, funded by Funda\c{c}\~{a}o de Amparo \`{a} Pesquisa do Estado de S\~{a}o Paulo, FAPESP, under the Research Grant No.\ 2013/07375-0\@. The authors further acknowledge the National Laboratory for Scientific Computing (LNCC/MCTI, Brazil) for providing HPC resources of the SDumont supercomputer. This work was also granted access to the HPC resources of IDRIS under the allocation 2020-A0092A12067 made by GENCI. The first author acknowledges authorization by his employer, Embraer S.A., which has allowed his participation in the present research effort. The doctoral scholarship provide by FAPESP to the third author, under the Grant No.\ 2018/05524-1\@, is thankfully acknowledged. Additional support to the fourth author under the FAPESP Research Grant No.\ 2013/07375-0\@ is also gratefully acknowledged.

\section{REFERENCES} 
\label{Sec:references}

\bibliographystyle{abcm}
\renewcommand{\refname}{}
\bibliography{bibfile_paper}

\section{RESPONSIBILITY NOTICE}

The authors are solely responsible for the printed material included in this paper.

\end{document}